%% file: resub.tex
\documentstyle[epic,eepic,epsf]{article}
\def\ssim{\mbox{\hspace{1ex}}
          \hbox{\raisebox{.4ex}{$<$}}\mbox{\hspace{-1.7ex}}
         {\lower.8ex \hbox{$\sim$}}\mbox{\hspace{1.1ex}}}
\def\gsim{\mbox{\hspace{1ex}}
          \hbox{\raisebox{.4ex}{$>$}}\mbox{\hspace{-1.7ex}}
         {\lower.8ex \hbox{$\sim$}}\mbox{\hspace{1.1ex}}}
\title{
Non-perturbative approach to the effective potential
of the $\lambda\phi^{4}$ theory at finite temperature
}

\author{
 Tomohiro Inagaki$^{a}$%
  \thanks{e-mail address: inagaki@hetsun1.phys.kobe-u.ac.jp},
 Kenzo Ogure$^{b}$%
  \thanks{e-mail address: ogure@icrr.u-tokyo.ac.jp} and
 Joe Sato$^{c}$%
  \thanks{e-mail address: joe@hep-th.phys.s.u-tokyo.ac.jp}
\\
  {\small
   {\it $^a$ Department of Physics, Kobe University,}
   }\\
  {\small
     {\it Rokkoudai, Nada, Kobe 657, Japan}
   }\\
   {\small 
     {\it $^b$Institute for Cosmic Ray Research, University of Tokyo,}
  }\\
  {\small {\it Midori-cho, Tanashi, Tokyo 188-8502, Japan } }
  \\
  {\small {\it $^c$ Department of Physics, University of Tokyo,} }\\
  {\small {\it Bunkyo-ku, Hongo, Tokyo 133-0033, Japan } }
}

\begin{document}

\maketitle

\begin{abstract}
We construct a non-perturbative method to 
calculate the effective potential 
of the $\lambda\phi^{4}$ 
theory at finite temperature.
We express the derivative of the effective potential with respect
to the mass square in terms of the full 
propagator. We reduce it to the partial 
differential equation for the effective 
potential using an approximation.
We numerically solve it
and obtain the effective potential non-perturbatively.
We find that the phase transition is second order as it should be.
We determine several critical exponents.
\\[3mm]
\end{abstract}
%\end{frontmatter}

\section{Introduction}\label{intro}

It is often expected that broken symmetries are restored
at high temperature \cite{KL}. 
The temperature-induced phase transition
will be observed in relativistic heavy ion collisions,
interior of neutron stars, and the early stage of the universe.
We may probe new physics through the phase transition 
at high temperature.

It is, however, very difficult to examine the phase transition.
For example, the perturbation theory often breaks down at high temperature.
As is well-known in finite temperature field theories
higher order contributions of the loop expansion are
enhanced for Bose fields by many interactions in the
thermal bath \cite{Sha,Arn3}.
In the $\lambda\phi^{4}$ theory physical quantities
are expanded in terms of
$\lambda T^{2}/m^{2}$ and 
$\lambda T/m$ at finite temperature.
The ordinary loop expansion is improved by resumming the
daisy diagram which includes all the higher order
contributions of 
$\displaystyle {\cal O}\left(\left(\lambda T^{2}/m^{2}
\right)^{n}\right)$ \cite{Car,Arn2,DJ,W,Tak,Fen,BFH,FH,K,Bell}
The loop expansion parameter is $\lambda T/m$ after the resummation. 
It means that the perturbation theory breaks down 
at $T \gsim m/\lambda$ \cite{Fen}.
Around the critical temperature the ratio $m/T$
is always of ${\cal O}(\lambda)$ so a non-perturbative 
analysis is necessary to study the phase transition in
$\lambda\phi^{4}$ theory\cite{Arn2}.

A variety of methods is used to investigate the phase transition, for 
example, lattice simulation \cite{FHJJM,CFHJJM,GIS,GIKPS,KLRS,Ao},
C.J.T. method \cite{CJT,AC},
$\varepsilon$-expansion \cite{AY},
effective three dimensional theory \cite{Gin,AP,Nad,Land,FKRS},
gap equation method \cite{BP}, non-purturbative renormalization group
method\cite{MT,DM,AMSST,ABBFTW,BW} and so on.
All the same we still need another method to
study the phase transition since they are applicable to limited
situations.

In Ref.\cite{DHL} a new non-perturbative
approach was suggested to avoid the 
infrared divergence which appears in the pressure \cite{Linde}. 
They differentiated the generating functional with respect 
to the mass square and found the infrared finite expression for the 
pressure in thermal equilibrium.

In the present paper we employ the idea of Ref.\cite{DHL} and develop
a new method to calculate the effective potential. Differentiating the
effective potential with respect to the mass square, we express the
derivative in terms of the full propagator. We construct the partial
differential equation for the effective potential by approximating the
full propagator. We calculate the effective potential beyond the
perturbation theory by solving this equation.

In section 2 we consider the $\lambda\phi^{4}$ theory at finite
temperature and show the exact expression of the derivative of the
effective potential $\displaystyle \frac{\partial V}{\partial m^{2}}$.
We approximate it and obtain the partial differential equation for the
effective potential.  We give the reasonable initial condition to
solve this equation.  In Sec.3 we solve it and get the effective
potential numerically.  We obtain the susceptibility, field
expectation value, and specific heat from it. We determine the several
critical exponents by observing their behaviours as $T$ varies.  The
Sec.4 is devoted to the concluding remarks.

\section{Evolution equation for the effective potential}

As mentioned in Sec.\ref{intro}, the loop expansion loses its
validity at high temperature. 
We need a non-perturbative method to calculate the effective
potential. 
The effective potential, in general, satisfies the following relation,
\begin{equation}
     V(m^{2})=\int^{m^{2}}_{M^{2}}\left(
     \frac{\partial V}{\partial m^{2}} 
     \right)dm^{2}
     +V(M^{2}).
\label{iden}
\end{equation}
Once we know $\displaystyle 
\frac{\partial V}{\partial m^{2}}$ and $V(M^{2})$, we
can calculate the effective potential for arbitrary $m^2$. Following the 
idea, we construct an evolution equation for the effective potential of
the $\lambda\phi^4$ theory at finite temperature. 
In the following we give $\displaystyle
\frac{\partial V}{\partial m^{2}}$ and an
appropriate initial condition $V(M^{2})$.

We consider the $\lambda\phi^{4}$ theory
which is defined by the Lagrangian density
\begin{equation}
     {\cal L}_{E}=-\frac{1}{2}
     \left(\frac{\partial \phi}{\partial \tau}\right)^{2}
     -\frac{1}{2}(\mbox{\boldmath $\nabla$}\phi)^{2}
     -\frac{1}{2}m^{2}\phi^{2}
     -\frac{\lambda}{4!}\phi^{4}
     +{\cal L}_{ct}
     +J\phi,
\label{lag}
\end{equation}
where ${\cal L}_{ct}$ represents the counter term and
$J$ is an external source function.
If $m^2$ is negative, the scalar field $\phi$ develops the
non-vanishing field expectation value at $T=0$.
It is expected that the field expectation value
 decreases as T increases and the phase transition
takes place at the critical temperature $T_{c}$. We can explore
properties of this phase transition by studying
the effective potential at finite temperature.

Following the standard procedure of dealing with the
Matsubara Green function \cite{Mas}, we introduce the temperature
to the theory. The generating functional
at finite temperature is given by
\begin{equation}
     Z_{T}=\int D[\phi]\exp \left(
     \int^{1/T}_{0}d\tau\int d^{3}
     \mbox{\boldmath $x$}
     {\cal L}_{E}\right).
\label{gfunc1}
\end{equation}
In the $\lambda\phi^{4}$ theory
the derivative of the effective potential
$\displaystyle \frac{\partial V}{\partial m^{2}}$
is expressed by the full propagator
of the scalar field (See Appendix A),
\begin{equation}
     \frac{\partial V}{\partial m^{2}}=
     \frac{\partial V_{tree}}{\partial m^{2}}
     +\frac{\partial V_{1}}{\partial m^{2}}
     +\frac{\partial V_{2}}{\partial m^{2}}
     +\frac{\partial V_{ct}}{\partial m^{2}}.
\label{delv3}
\end{equation}
$V_{tree}$ is the tree part, 
\begin{equation}
     \frac{\partial V_{tree}}{\partial m^{2}}\equiv
     \frac{1}{2}\bar{\phi}^{2} .
\label{delvtree}
\end{equation}
The non-perturbative effects are contained in $V_{1}$ and $V_{2}$,
\begin{eqnarray}
     \frac{\partial V_{1}}{\partial m^{2}}&\equiv &
     \frac{1}{2\pi i}
     \int^{+i\infty +\epsilon}_{-i\infty +\epsilon}dp_{0}
     \int\frac{d^{3}\mbox{\boldmath $p$}}{(2\pi)^{3}}
     \frac{1}{-p_{0}^{2}+\mbox{\boldmath $p$}^{2}+m^{2}
     +\frac{\lambda}{2}
     \bar{\phi}^{2}+\Pi}\frac{1}{e^{p_{0}/T}-1}\\
\label{delv31}
     \frac{\partial V_{2}}{\partial m^{2}}&\equiv &
     \frac{1}{4\pi i}\int^{+i\infty}_{-i\infty}dp_{0}
     \int\frac{d^{3}\mbox{\boldmath $p$}}{(2\pi)^{3}}
     \frac{1}{-p_{0}^{2}+\mbox{\boldmath $p$}^{2}+m^{2}
     +\frac{\lambda}{2}\bar{\phi}^{2}+\Pi} .
\label{delv32}
\end{eqnarray}
$V_{ct}$ is the counter term part,
\begin{eqnarray}
     \frac{\partial V_{ct}}{\partial m^{2}}&\equiv &
     (Z_{m}Z_{\phi}-1)\biggl[\frac{1}{2}\bar{\phi}^{2}
     \nonumber \\
     &&+\frac{1}{2\pi i}
     \int^{+i\infty +\epsilon}_{-i\infty +\epsilon}dp_{0}
     \int\frac{d^{3}\mbox{\boldmath $p$}}{(2\pi)^{3}}
     \frac{1}{-p_{0}^{2}+\mbox{\boldmath $p$}^{2}+m^{2}
     +\frac{\lambda}{2}
     \bar{\phi}^{2}+\Pi}\frac{1}{e^{p_{0}/T}-1}\nonumber \\
     &&+\frac{1}{4\pi i}\int^{+i\infty}_{-i\infty}dp_{0}
     \int\frac{d^{3}\mbox{\boldmath $p$}}{(2\pi)^{3}}
     \frac{1}{-p_{0}^{2}+\mbox{\boldmath $p$}^{2}+m^{2}
     +\frac{\lambda}{2}\bar{\phi}^{2}+\Pi}\biggr].
\label{delv3ct}
\end{eqnarray}
Here $\Pi=\Pi(\mbox{\boldmath $p$}^{2},-p_{0}^{2},\bar{\phi},m^{2},T)$ 
describes the full self-energy.
The third term $\displaystyle \frac{\partial V_{2}}{\partial m^{2}}$ 
on the right hand side of Eq.(\ref{delv3}) is 
divergent. This divergence is removed by the counter term
(\ref{delv3ct}) after the usual renormalization 
procedure is adopted at $T=0$.
The counter term which is determined at $T=0$ removes
the ultra-violet divergence even in finite temperature \cite{W,K,Kis}.

We give the initial condition at $M^{2}\sim {\cal
O}(T^{2})$ where the loop 
expansion is valid, ($\lambda T/M \sim \lambda \ll 1$).
We calculate $V(M^2)$ by the
perturbation theory up to the one loop order.
After the renormalization with $\overline{\mbox{MS}}$ scheme 
at the renormalization scale $\bar{\mu}$, 
the one-loop effective potential is
\begin{equation}
     V(M^{2})=V_{tree}(M^{2})+V_{1}(M^{2})+V_{2}(M^{2})+V_{ct}(M^{2}),
\label{shoki}
\end{equation}
where $V_{tree}, V_{1}$ and $V_{2}+V_{ct}$ are given by
\begin{eqnarray}
     &&V_{tree}(M^{2})=\frac{1}{2}M^{2}\bar{\phi}^{2}
     +\frac{\lambda}{4!}\bar{\phi}^{4} ,
\label{shoki1}\\
     &&V_{1}(M^{2})=\frac{T}{2\pi^{2}}
     \int^{\infty}_{0}dr r^{2}\log \left[
     1-\exp\left(-\frac{1}{T}\sqrt{r^{2}+M^{2}+
     \frac{\lambda}{2}\bar\phi^{2}}\right)\right],
\label{shoki2}\\
     &&V_{2}(M^{2})+V_{ct}(M^{2})
     =\frac{\left(M^{2}+\frac{\lambda}{2}\bar\phi^{2}\right)^{2}}
     {64\pi^{2}}\left[\log\left(\frac{M^{2}+\frac{\lambda}{2}\bar\phi^{2}}
     {\bar{\mu}^{2}}\right)-\frac{3}{2}\right].
\label{shoki3}
\end{eqnarray}
Note that we need not resum the daisy diagram 
which has only a negligible contribution of 
${\cal O} (\lambda)$ for $M^{2} \sim {\cal O}(T^{2})$.

In order to investigate the 
temperature-induced phase transition we consider 
the theory with non-vanishing field expectation value 
at $T=0$ 
(i.e. $m^{2}$ takes a negative value, $m^{2}=-\mu^{2}$). 
We calculate $V(-\mu^{2})$ with the effective potential (\ref{shoki}) by
\begin{eqnarray}
     V(-\mu^{2})&=&\int^{-\mu^{2}}_{M^{2}}\left(
     \frac{\partial V_{tree}}{\partial m^{2}} 
     +\frac{\partial V_{1}}{\partial m^{2}} 
     +\frac{\partial V_{2}}{\partial m^{2}} 
     +\frac{\partial V_{ct}}{\partial m^{2}} 
     \right)dm^{2} \nonumber \\ 
     &&+V_{tree}(M^{2})
     +V_{1}(M^{2})
     +V_{2}(M^{2})
     +V_{ct}(M^{2}) .
\end{eqnarray}
For $m^{2} \ll T^{2}$ the contribution from
$\displaystyle \frac{\partial V_{1}}{\partial m^{2}}$ 
is enhanced by the Bose factor.
The contribution from $V_{1}$ can be the same order
as that from the tree part around the critical temperature.

The quantity $V_{2}+V_{ct}$ will have a negligible 
contribution:
\begin{eqnarray}
     &&\int^{-\mu^{2}}_{M^{2}}\left(
     \frac{\partial V_{2}}{\partial m^{2}} 
     +\frac{\partial V_{ct}}{\partial m^{2}} 
     \right)dm^{2}
     +V_{2}(M^{2})
     +V_{ct}(M^{2})\nonumber \\ 
     &&=V_{2}(-\mu^{2})
     +V_{ct}(-\mu^{2}) .
\label{neglect}
\end{eqnarray}
We can show that 
$V_{2}(-\mu^{2})+V_{ct}(-\mu^{2})$ is really small
at the leading order of the loop expansion.
At one loop level with daisy diagram resummation
we find 
\begin{eqnarray}
     V_{2}(-\mu^{2})&+&V_{ct}(-\mu^{2})\nonumber \\
     &&=\frac{\left(-\mu^{2}+\frac{\lambda}{2}\bar\phi^{2}+\Pi\right)^{2}}
     {64\pi^{2}}\left[\log\left(\frac{-\mu^{2}
     +\frac{\lambda}{2}\bar\phi^{2}+\Pi}
     {\bar{\mu}^{2}}\right)-\frac{3}{2}\right].
\label{neglect1}
\end{eqnarray}
The self-energy satisfies $\Pi\sim \mu^{2}$
around the critical temperature for the second-order or
the weakly first-order phase transition \cite{Arn2}.
Because we are interested in the effective potential at small $\bar\phi$
region only to 
investigate the phase structure, we neglect (\ref{neglect}) in the 
following calculations.

Furthermore, we ignore the momentum dependence of the self-energy
$\Pi$ and generate the two point function from the effective potential
$V$. We replace as follows in Eq.(\ref{delv3}),
\begin{equation}
     m^{2}+\frac{\lambda}{2}\bar{\phi}^{2}+
     \Pi(0,0,\bar{\phi},m^{2},T)\rightarrow
     \frac{\partial^{2}V}{\partial\bar\phi^{2}}.
\label{okikae}
\end{equation}
We obtain the 
partial differential equation for the effective potential by integrating
over $p_{0}$ and angle variables in Eq.(\ref{delv3}), 
\begin{equation}
     \frac{\partial V}{\partial m^{2}}=
     \frac{1}{2}\bar{\phi}^{2}+\frac{1}{4\pi^{2}}
     \int^{\infty}_{0}dr r^{2}\frac{1}{\displaystyle \sqrt{r^{2}
     +\frac{\partial^{2}V}{\partial\bar\phi^{2}}}}
     \frac{1}{\displaystyle \exp\left(\frac{1}{T}\sqrt{r^{2}
     +\frac{\partial^{2}V}{\partial\bar\phi^{2}}}\right)-1}.
\label{partial}
\end{equation}

\section{Numerical results}

We calculate the effective potential by
solving the partial differential equation (\ref{partial})
with the initial condition $V_{tree}+V_{1}$ 
in Eq.(\ref{shoki}).
We solve the equation numerically and show the phase 
structure of $\lambda\phi^{4}$ theory.

\subsection{analytic continuation}

The integral in (\ref{partial}) is well defined in the region 
where $\displaystyle \frac{\partial^{2}V}{\partial{\bar \phi^{2}}}$
is real and positive.
The effective 
potential $V(\bar{\phi})$ is, however, complex for small $\bar{\phi}$
below the critical temperature, $T <  T_{c}$.  
We have to find the analytic continuation in order to calculate 
the effective potential there.

To make the analytic continuation, we change the variable of
integration $r$ to $z$ through 
\begin{equation}
     z=\sqrt{\frac{r^{2}}{T^{2}}+Z^{2}}-Z ,
    \label{cv2}
\end{equation}
and rewrite the differential equation (\ref{partial}),
\begin{equation}
     \frac{\partial V}{\partial m^{2}}=
     \frac{1}{2}\bar\phi^{2}+\frac{T^{2}}{4\pi^{2}}
     \int^{\infty}_{0}dz
     \frac{\sqrt{z(z+2Z)}}{e^{z+Z}-1}.
    \label{partial2}  
\end{equation}
Here $Z$ is the double valued function which is given by
$ \displaystyle Z=\sqrt{\frac{1}{T^{2}}\frac{\partial^{2} V}
{\partial \bar \phi^{2}}}$.

The imaginary part of the effective potential is interpreted as a decay
rate of the unstable state \cite{WW}.
It is natural that we assume
such an imaginary part is negative.
The imaginary part of $\displaystyle \frac{\partial V}{\partial m^{2}}$ 
should be positive in order that the imaginary part of the effective
potential may be negative.
We have to select the branch of 
$ \displaystyle Z=\sqrt{\frac{1}{T^{2}}\frac{\partial^{2} V}
{\partial \bar \phi^{2}}}$ so that imaginary part of 
$\displaystyle \frac{\partial V}{\partial m^{2}}$ will be positive.
We calculate the effective potential in this branch and the
imaginary part of it is always negative as we will see in the 
next subsection.

\subsection{numerical result}

Putting the initial condition $V_{tree}+V_{1}$ 
in Eq.(\ref{shoki}) at $M^{2}=T^{2}$ 
we numerically solve Eq.(\ref{partial2}) and obtain
the effective potential at $m^{2}=-\mu^{2}$.
We use the explicit 
differencing method \cite{NR}.
In this subsection we show the effective potential 
and calculate critical exponents.

\begin{figure}
    \begin{minipage}{.49\linewidth}
    \begin{center}
    \input{epot1_eepic.tex}
    (a) non-perturbative method
    \end{center}
    \end{minipage}
\hfill
    \begin{minipage}{.49\linewidth}
    \begin{center}
    \input{epotper1_eepic.tex}
    (b) perturbation at 1-loop level
    \end{center}
    \end{minipage}
\vglue 6ex
    \begin{minipage}{.49\linewidth}
    \begin{center}
    \input{epotper1p_eepic.tex}
    (b') perturbation at 1-loop level around the critical temperature
    \end{center}
    \end{minipage}
\hfill
    \begin{minipage}{.49\linewidth}
    \begin{center}
    \input{epotper2_eepic.tex}
    (c) perturbation at 2-loop level
    \end{center}
    \end{minipage}
\vglue 2ex
\caption{The behaviour of the effective potential $V$
is shown for fixed $\lambda (=1)$ as the function of the 
temperature. We find no qualitative change for other values
$\lambda(=0.5,0.1,0.05)$. We normalise that $V(0)=0$.}
\label{effpot}
\end{figure}
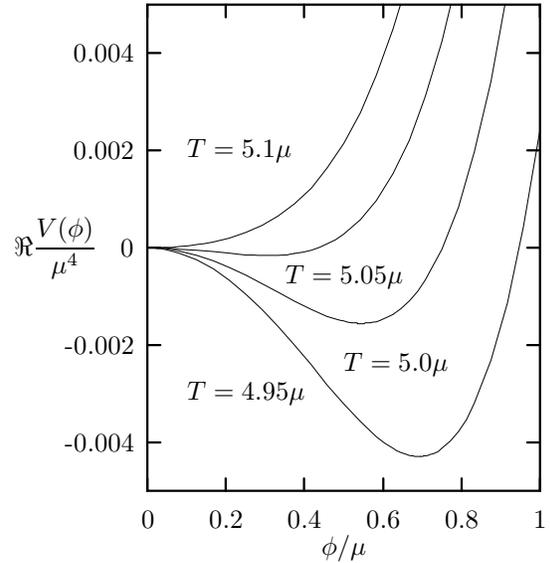
We illustrate the behaviour of the effective potential at $\lambda =1$
in Fig.\ref{effpot} (a).  The field expectation value $\phi_{c}$ is
the minimum point of the effective potential. It seems to disappear
smoothly at the critical temperature.  We find that the phase
transition is second order as it should be.

For comparison, in Figs.\ref{effpot} (b), (b') and (c) we show the
effective potential calculated by the perturbation theory at one and
two loop order with daisy diagram resummation.\footnote{ We use the
  equations in Ref\cite{Arn2} to draw them.  }At the one loop order an
extremely small gap appears at the critical temperature as is clearly
seen in Fig.\ref{effpot} (b').  The phase transition is first order at
the one loop order.

\begin{figure}
\unitlength=1cm
\begin{picture}(16,4)
%\input zahyou
%\zahyou{16}{6}
\unitlength=1mm
%\put(25,80){$f(\Delta_{31},\epsilon = 0.03)$}
%\put(150,5){$\Delta_{31}$}
\centerline{
\epsfxsize=13cm
\epsfbox{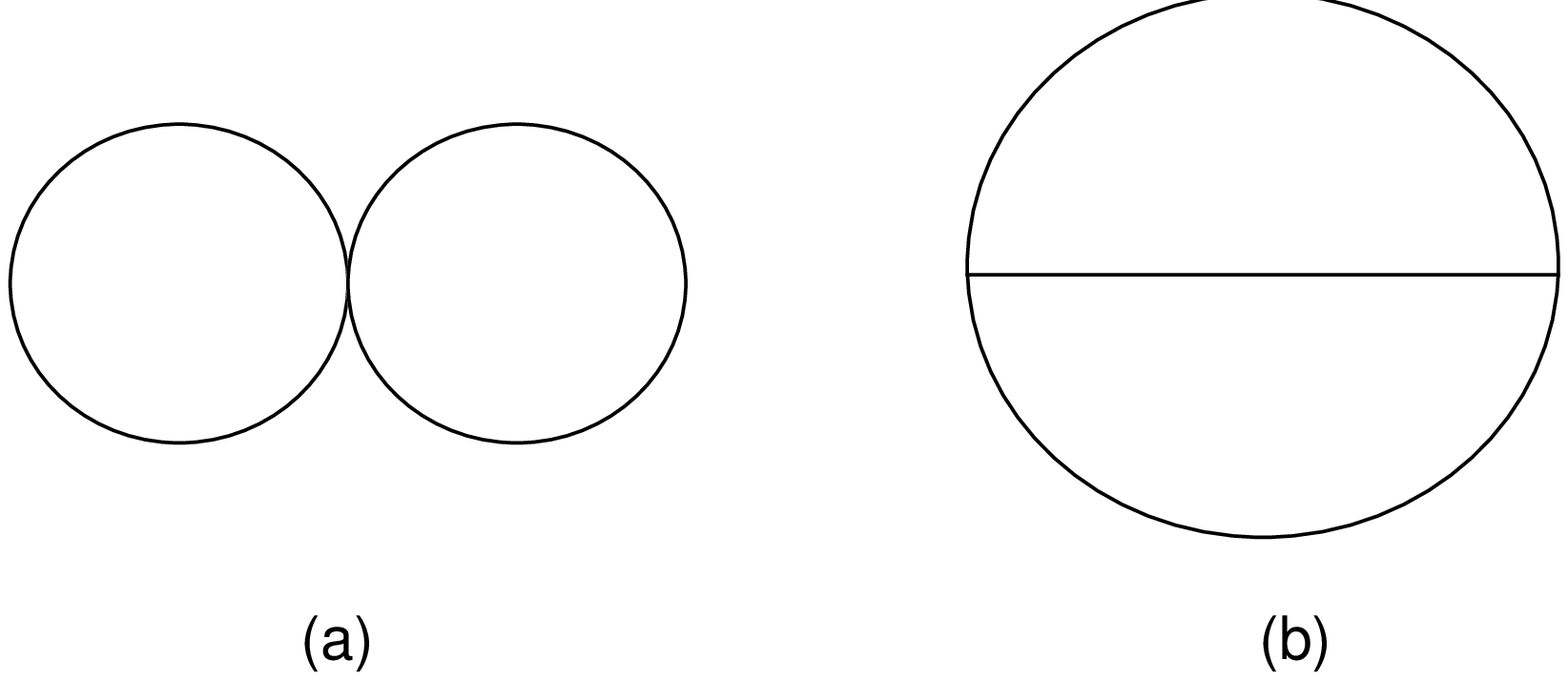}
}
\end{picture}
\caption{
Two loop diagrams that contribute to the effective potential.
}
\label{fig8}
\end{figure}

This situation is modified at the two loop order.  We observe no gap
and find that the phase transition is second order as shown in
Fig.\ref{effpot} (c).  Though Fig.\ref{effpot} (a) and
Fig.\ref{effpot} (c) show the similar behaviour, it will be accident.
The effective potential calculated up to the two loop order includes
the contribution from the graphs, Fig.\ref{fig8} (a) and
Fig.\ref{fig8} (b), with daisy resummation.  On the other hand we can
take into account the contribution from all the other graphs in
addition to Fig.\ref{fig8} (a) and Fig.\ref{fig8} (b) within the
approximation (\ref{okikae}) by solving Eq.(\ref{partial2})
automatically. The Fig.\ref{effpot} (a) accidentally coincides with
Fig.\ref{effpot} (c).

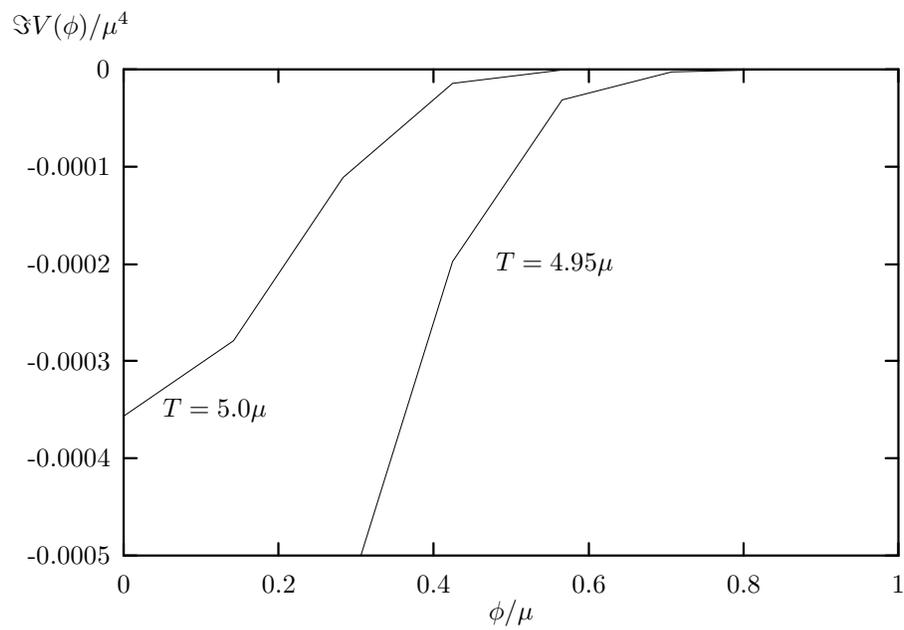
\begin{figure}
    \begin{center}
    \input{epot2_eepic.tex}
    \end{center}
\caption{Imaginary part of the effective potential near the 
critical temperature}
\label{imaeffpot}
\end{figure}
For $T < T_{c}$ the effective potential develops a non-vanishing
imaginary part at small $\bar\phi$ range. We show it in Fig.\ref{imaeffpot}.
It should be noted that the sign of the imaginary part is always negative. 
It is consistent with the discussion
in the previous subsection.

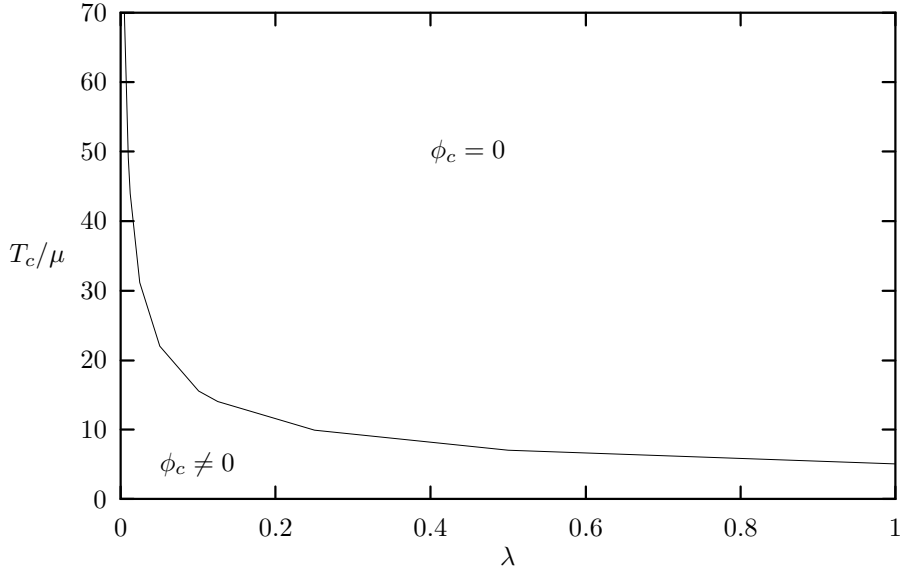
\begin{figure}
\begin{center}
\input{cri_eepic.tex}
\end{center}
\caption{Phase boundary}
\label{phst}
\end{figure}
Evaluating the effective potential with varying the temperature, $T$,
and the coupling constant, $\lambda$, we obtain the
critical temperature as a function of $\lambda$ where the field
expectation value disappears. 
We show the phase boundary on $T$-$\lambda$ plain
in Fig.\ref{phst}.

The critical exponents
are defined for the second-order phase transition.
Around the critical temperature we expect that the 
susceptibility $\chi$, the expectation value $\phi_{c}$, and 
the specific heat $C$ behave as \cite{Pes}
\begin{equation}
     \chi\propto |t|^{-\gamma},
     \phi_{c}\propto |t|^{+\beta},
     C\propto |t|^{-\alpha},
    \label{criex}
\end{equation}
where $t =(T-T_{c})/T$.
Analysing the effective potential more precisely we can calculate
the critical exponents $\gamma, \beta$, and $\alpha$.
The susceptibility $\chi$ satisfies the following relation,
\begin{equation}
     \xi\propto\rho^{-1},
    \label{xi}
\end{equation}
where $\rho$ is the curvature of the effective potential
at $\phi_{c}$.

Since the specific heat $C$ is given by the second 
derivative of the effective potential around the critical 
temperature,  the effective potential $V(\phi_{c})$
behaves as
\begin{equation}
     V(\phi_{c})\propto |t|^{2-\alpha}.
    \label{c}
\end{equation}

\begin{figure}
    \begin{minipage}{.49\linewidth}
    \begin{center}
    \input{criphi_eepic.tex}
    (a) Field expectation value $\phi_{c}$
    \end{center}
    \end{minipage}
\hfill
    \begin{minipage}{.49\linewidth}
    \begin{center}
    \input{crirho_eepic.tex}
    (b) curvature $\rho$ at the minimum
    \end{center}
    \end{minipage}
\vglue 6ex
    \begin{minipage}{.49\linewidth}
    \begin{center}
    \input{criv_eepic.tex}
    (c) Minimum of the effective potential $V(\phi_{c})$
    \end{center}
    \end{minipage}
\vglue 2ex
\caption{Critical behaviour of $\phi_{c}$, $\rho$ and
$V(\phi_{c})$}
\label{cribeh}
\end{figure}
We examine the behaviour of $\rho, \phi_{c}$
,and $V(\phi_{c})$ around the critical temperature
and find the critical exponents $\gamma, \beta$ and $\alpha$.
In Fig.$\ref{cribeh}$ the critical behaviour of 
$\rho, \phi_{c}$ and $V(\phi_{c})$ are shown as a function
of the temperature.
We numerically calculate the critical exponents from them.
Our numerical results are presented in 
the table \ref{results}.\footnote{
Due to the instability in the explicit 
differencing method, we can not see the fine structure of the
effective potential and can get only the rough values of the
critical exponents. We need further numerical study to get more
precise values. } 
\begin{table}[ht]
\caption{Critical exponents}
\begin{tabular}{|c|c|c|c|} 
     \hline
     & \makebox[40mm]{our results}
     & \makebox[40mm]{Landau theory} 
     & \makebox[40mm]{experimental results \cite{ZJ}}  \\
     \hline
     $\beta$  & $\sim 0.5$ & $0.5$ & $0.33$ \\
     \hline
     $\gamma$    & $\sim 1$ & $1$ & $1.24$ \\
     \hline
     $\alpha$ &  $\sim 0$  &  $0$  & $0.11$ \\
     \hline
\end{tabular}
\label{results}
\end{table}
\noindent
The critical exponents within our approximation are
independent of the coupling constant $\lambda$.
We note that our results described in the present 
subsection remain unchanged even when we put the 
initial mass scale $M^2=T^2/4$ or $M^2=4 T^2$.

\section{Conclusion}

We constructed the non-perturbative method to 
investigate the phase structure of $\lambda \phi^{4}$ theory.
The derivative of the effective potential with respect to 
mass square was exactly expressed in terms of the full propagator
at finite temperature.
We found the partial differential equation for the effective potential
with the replacement (\ref{okikae}).
We gave the initial condition by the 1 loop effective
potential in the range where the perturbation theory 
is reliable. We numerically solved the partial differential
equation and obtained the effective potential.

Though we made the approximation (\ref{okikae}),
we could find that 
the phase transition of $\lambda \phi^4$ theory 
is second order as it should be.
Our method is very interesting because it can show
the correct order of the phase transition.
The approximation (\ref{okikae}) may be fairy good.

We determined several critical exponents which roughly agree with
those of Landau approximation. They are, however, rough 
values because it is very difficult to solve the nonlinear
partial differential equation (\ref{partial}) numerically.
We need elaborate a numerical study to obtain more accurate critical
exponents.

The main problem of our non-perturbative method is how to 
improve the approximation to the full propagator. 
We cannot estimate the error from the approximation (\ref{okikae}).
We need improve the approximation to the full propagator 
in order to know the correction to the current result.

Our method is very promising since it can probe the region
where the traditional perturbation theory breaks down.

\section*{Acknowledgements}

The authors would like to thank Akira Niegawa and Jiro Arafune
for useful discussions.

\appendix
\section{The derivative of the effective potential in terms of
the full propagator}

The derivative of the effective potential
$\displaystyle \frac{\partial V}{\partial m^{2}}$ 
can be represented by the full propagator.
In this appendix we present details of
the calculation of
$\displaystyle \frac{\partial V}{\partial m^{2}}$
given in Eq.(\ref{delv3}).

We consider the Lagrangian density which is
defined by
\begin{equation}
     {\cal L}_{E}=-\frac{1}{2}
     \left(\frac{\partial \phi_{0}}{\partial \tau}\right)^{2}
     -\frac{1}{2}(\mbox{\boldmath $\nabla$}\phi_{0})^{2}
     -\frac{1}{2}m^{2}_{0}\phi^{2}_{0}
     -\frac{\lambda_{0}}{4!}\phi_{0}^{4}
     +J_{0}\phi_{0},
\label{lag0}
\end{equation}
where the suffix $0$ denotes the bare quantities.

We adopt the mass-independent renormalization
procedure and
represent the effective potential as a function of 
renormalized quantities.
The renormalization constants $Z$ and 
renormalized quantities are introduced through
transformations
\begin{equation}
\begin{array}{rcl}
     \phi_{0}&=&Z_{\phi}^{1/2}\phi ,\\
     m_{0}&=&Z_{m}^{1/2}m ,\\
     \lambda_{0}&=&Z_{\lambda}\lambda ,\\
     J_{0}&=&Z_{\phi}^{-1/2}J .
\end{array}
\label{ren}
\end{equation}
Using these renormalization constants and 
renormalized quantities, we separate the Lagrangian
density (\ref{lag0}) into the tree part ${\cal L}_{1}$
and the counter term part ${\cal L}_{ct}$ as \cite{Pes,Jac}
\begin{equation}
     {\cal L}_{E}={\cal L}_{1}[\phi]
     +{\cal L}_{ct}[\phi]+(J_{1}+J_{ct})\phi  ,
\label{div}
\end{equation}
where $J_{1}+J_{ct}\equiv J$.  The laglangian density ${\cal L}_{1}$
and ${\cal L}_{ct}$ are given by
\begin{equation}
\left\{
\begin{array}{rcl}
     {\cal L}_{1}[\phi]&\equiv&\displaystyle -\frac{1}{2}
     \left(\frac{\partial\phi}{\partial\tau}\right)^{2}
     -\frac{1}{2}(\mbox{\boldmath $\nabla$}\phi)^{2}
     -\frac{1}{2}m^{2}\phi^{2}-\frac{\lambda}{4!}\phi^{4},\\
     {\cal L}_{ct}[\phi]&\equiv&\displaystyle -\frac{1}{2}
     (Z_{\phi}-1)\left[
     \left(\frac{\partial\phi}{\partial\tau}\right)^{2}
     +(\mbox{\boldmath $\nabla$}\phi)^{2}\right]
     -\frac{1}{2}(Z_{m}Z_{\phi}-1)m^{2}\phi^{2}\\
     &&\displaystyle 
     -\frac{\lambda}{4!}(Z_{\lambda}Z_{\phi}^{2}-1)\phi^{4}.
\end{array}
\right.
\label{new L}
\end{equation}
Here we separate the external source $J$ into $J_{1}$ and 
$J_{ct}$, which satisfy the following equations:
\begin{equation}
\left\{
\begin{array}{rcl}
     \displaystyle \left.
     \frac{\partial{\cal L}_{1}}{\partial \phi}\right|_{\phi=\bar{\phi}}
+J_{1}&=&0 ,\\
     \langle \phi \rangle_{J}&=&\bar{\phi} . 
\end{array}
\right.
\label{ext1}
\end{equation}
We expand the field $\phi(x)$ around the classical
background $\bar{\phi}$,
\begin{equation}
     \phi(x)=\bar{\phi}+\eta(x),
\label{henkan}
\end{equation}
and then $\eta (x)$ satisfies
\begin{equation}
     \langle \eta \rangle_{J}=0 .
\label{barphieta}
\end{equation}

In terms of $\bar{\phi}$ and $\eta(x)$ Eq.(\ref{new L})
is rewritten as
\begin{eqnarray}
     {\cal L}_{1}+J_{1}\phi&=&
     -\frac{1}{2}m^{2}\bar{\phi}^{2}
     -\frac{\lambda}{4!}\bar{\phi}^{4}
     +J_{1}\bar{\phi}
     -\frac{1}{2}\left[
     \left(\frac{\partial\eta}{\partial\tau}\right)^{2}
     +(\mbox{\boldmath $\nabla$}\eta)^{2}\right]\nonumber \\
     &&-\frac{1}{2}\left(m^{2}
     +\frac{\lambda}{2}\bar{\phi}^{2}\right)\eta^{2}
     -\frac{\lambda}{3!}\bar{\phi}\eta^{3}
     -\frac{\lambda}{4!}\eta^{4}
      \nonumber \\
     &\equiv&{\cal L}_{1}(\bar{\phi})+{\cal L}'_{1}[\eta],
\label{renlag}
\end{eqnarray}
and
\begin{eqnarray}
     {\cal L}_{ct}+J_{ct}\phi&=&
     -\frac{1}{2}(Z_{m}Z_{\phi}-1)m^{2}\bar{\phi}^{2}
     -\frac{\lambda}{4!}(Z_{\lambda}Z_{\phi}^{2}-1)
     \bar{\phi}^{4}
     +J_{ct}\bar{\phi}\nonumber \\
     &&\left[-(Z_{m}Z_{\phi}-1)m^{2}\bar{\phi}
     -\frac{\lambda}{3!}(Z_{\lambda}Z_{\phi}^{2}-1)
     \bar{\phi}^{3}+J_{ct}\right]\eta\nonumber \\
     &&-\frac{1}{2}(Z_{\phi}-1)\left[
     \left(\frac{\partial\eta}{\partial\tau}\right)^{2}
     +(\mbox{\boldmath $\nabla$}\eta)^{2}\right]
     -\frac{1}{2}(Z_{\phi}Z_{m}-1)m^{2}\eta^{2}\nonumber\\
     &&-\frac{\lambda}{2}(Z_{\lambda}Z_{\phi}^{2}-1)
     \bar{\phi}^2\eta^{2}-\frac{\lambda}{3!}(Z_{\lambda}Z_{\phi}^{2}-1)
     \bar{\phi}\eta^{3}
     -\frac{\lambda}{4!}(Z_{\lambda}Z_{\phi}^{2}-1)\eta^{4}
     \nonumber \\
     &\equiv&{\cal L}_{ct}(\bar{\phi})+K\eta
     +{\cal L}'_{ct}[\eta],
\label{coutlag}
\end{eqnarray}
where $K$ is defined by
\begin{equation}
     K\equiv -(Z_{m}Z_{\phi}-1)m^{2}\bar{\phi}
     -\frac{\lambda}{3!}(Z_{\lambda}Z_{\phi}^{2}-1)\bar{\phi}^{3}
     +J_{ct}.
\label{def:K}
\end{equation}
From Eq.(\ref{barphieta}),  $\eta (x)$ satisfies
\begin{equation}
\langle\eta \rangle_{K}=0.
\label{etak}
\end{equation}

The generating functional $W_{T}[J]$ for connected
Green functions is given by
\begin{equation}
     Z_{T}[J]=e^{W_{T}[J]}.
\label{gfunc2}
\end{equation}
The effective action $\Gamma_{T}(\bar{\phi})$
is defined as the Legendre transformation of $W_{T}[J]$.
In the spacetime with the translational invariance
the effective potential $V(\bar{\phi})$ is
proportional to the effective action.
The effective potential $V(\bar{\phi})$ is
\begin{equation}
     -\frac{\Omega}{T}V({\bar \phi})=\Gamma_{T}(\bar{\phi})
     =W_{T}[J]-\frac{\Omega}{T}\bar{\phi}J,
\label{effact}
\end{equation}
where $\displaystyle \Omega=\int d^{3}\mbox{\boldmath $x$}$
and the new variable $\bar{\phi}$ is given by
\begin{equation}
     \frac{\delta}{\delta J(y)}W_{T}[J]
     =\bar{\phi}(y)=\bar{\phi}=\mbox{const}.
\label{barphi}
\end{equation}
Substituting Eqs.(\ref{renlag}) and (\ref{coutlag}) 
into Eq.(\ref{div}), we rewrite the generating functional $Z_{T}$ 
 as a functional of renormalized quantities,
\begin{eqnarray}
     Z_{T}&=&\exp\left\{\frac{\Omega}{T}
     \left[{\cal L}_{1}(\bar{\phi})+{\cal L}_{ct}(\bar{\phi})
     +(J_{1}+J_{ct})\bar{\phi}\right]\right\}\nonumber\\
     &&\times\int D[\eta]\exp\int^{1/T}_{0}d\tau
     \int d^{3}\mbox{\boldmath $x$}
     \left({\cal L}'_{1}[\eta]+{\cal L}'_{ct}[\eta]
     +K\eta\right)\nonumber \\
     &\equiv&\exp\left\{\frac{\Omega}{T}
     \left[{\cal L}_{1}(\bar{\phi})+{\cal L}_{ct}(\bar{\phi})
     +(J_{1}+J_{ct})\bar{\phi}\right]\right\}Z'(K).
\label{gfunc3}
\end{eqnarray}
Taking into account Eqs.(\ref{gfunc2}) and (\ref{effact}), 
we obtain the effective potential from Eq.(\ref{gfunc3}):
\begin{equation}
     V(\bar{\phi})=-
     \left[{\cal L}_{1}(\bar{\phi})+{\cal L}_{ct}(\bar{\phi})
     \right]-\frac{T}{\Omega}\log Z'(K).
\label{effact2}
\end{equation}
In the mass-independent renormalization procedure
the renormalization constants $Z$ are independent
of the mass $m$. We easily differentiate the effective
potential $V$ by the mass square $m^{2}$ and get
\begin{eqnarray}
     \frac{\partial V(\bar{\phi})}{\partial m^{2}}
     &=&\left[\frac{1}{2}\bar{\phi}^{2}
     +\frac{1}{2}(Z_{m}Z_{\phi}-1)\bar{\phi}^{2}\right]
     \nonumber \\
     &&-\frac{T}{\Omega}\frac{1}{Z'}\int D[\eta]
     \left[-\frac{1}{2}\eta^{2}
     +\frac{\partial K}{\partial m^{2}}\eta
     -\frac{1}{2}(Z_{m}Z_{\phi}-1)\eta^{2}\right]\nonumber \\
     &&\times\exp\int^{1/T}_{0} d\tau\int d^{3}
     \mbox{\boldmath $x$}\left({\cal L}'_{1}[\eta]+{\cal L}'_{ct}[\eta]
     +K\eta \right)
\label{delv}
     \\
     &=&\left[\frac{1}{2}\bar{\phi}^{2}
     +\frac{1}{2}(Z_{m}Z_{\phi}-1)\bar{\phi}^{2}
     +\frac{1}{2}\langle\eta^{2}(0)\rangle
     +\frac{1}{2}(Z_{m}Z_{\phi}-1)\langle\eta^{2}(0)\rangle\right],
     \nonumber
\end{eqnarray}
where we use Eq.(\ref{etak}) \cite{DHL}.

For $\lambda\phi^{4}$ theory the two-point function
$\langle\eta^{2}(0)\rangle$ in Eq.(\ref{delv}) is rewritten as
\begin{eqnarray}
     \langle\eta^{2}(0)\rangle &=&\int^{1/T}_{0} d\tau\int d^{3}
     \mbox{\boldmath $x$}
     \delta^{3}(\mbox{\boldmath $x$}) \delta_{T}(\tau)
     \langle\eta(\mbox{\boldmath $x$},\tau)
     \eta(\mbox{\boldmath $0$},0)\rangle \nonumber \\
     &=&T\sum^{\infty}_{n=-\infty}
     \int\frac{d^{3}\mbox{\boldmath $p$}}{(2\pi)^{3}}
     \int^{1/T}_{0} d\tau\int d^{3}
     \mbox{\boldmath $x$}e^{i(\mbox{\boldmath $xp$}+\omega_{n}\tau)}
     \langle\eta(\mbox{\boldmath $x$},\tau)
     \eta(\mbox{\boldmath $0$},0)\rangle
\label{comp}
     \\
     &=&T\sum^{\infty}_{n=-\infty}
     \int\frac{d^{3}\mbox{\boldmath $p$}}{(2\pi)^{3}}
     \frac{1}{\omega_{n}^{2}+\mbox{\boldmath $p$}^{2}+m^{2}
     +\frac{\lambda}{2}\bar{\phi}^{2}+\Pi},
     \nonumber
\end{eqnarray}
where $\omega_{n}=2\pi nT$ due to the periodic boundary
condition for Bose fields and 
$\Pi=\Pi(\mbox{\boldmath $p$}^{2},-p_{0}^{2},\bar{\phi},m^{2},T)$ 
is the full self-energy.
Substituting Eq.(\ref{comp}) into Eq.(\ref{delv}), we express the
derivative of the effective potential in terms of the full propagator,
\begin{eqnarray}
     &&\frac{\partial V}{\partial m^{2}}=
     \frac{1}{2}\bar{\phi}^{2}+\frac{T}{2}
     \sum^{\infty}_{n=-\infty}
     \int\frac{d^{3}\mbox{\boldmath $p$}}{(2\pi)^{3}}
     \frac{1}{\omega_{n}^{2}+\mbox{\boldmath $p$}^{2}+m^{2}
     +\frac{\lambda}{2}\bar{\phi}^{2}+\Pi} 
\label{delv2}
     \\
     &&+(Z_{m}Z_{\phi}-1)\left[\frac{1}{2}\bar{\phi}^{2}
     +\frac{T}{2}\sum^{\infty}_{n=-\infty}
     \int\frac{d^{3}\mbox{\boldmath $p$}}{(2\pi)^{3}}
     \frac{1}{\omega_{n}^{2}+\mbox{\boldmath $p$}^{2}+m^{2}
     +\frac{\lambda}{2}\bar{\phi}^{2}+\Pi}\right].
     \nonumber
\end{eqnarray}
Using the residue theorem, we convert the frequency 
sum $\displaystyle T\sum^{\infty}_{n=-\infty}$
to contour integrals.
As long as $\Pi$ has no singularity along the 
imaginary $p_{0}$ axis, Eq.(\ref{delv2}) naturally separates
into a piece which contains a Bose factor and a piece which
does not \cite{K,Kis},
\begin{eqnarray}
     \frac{\partial V}{\partial m^{2}}&=&
     \frac{1}{2}\bar{\phi}^{2}+\frac{1}{2\pi i}
     \int^{+i\infty +\epsilon}_{-i\infty +\epsilon}dp_{0}
     \int\frac{d^{3}\mbox{\boldmath $p$}}{(2\pi)^{3}}
     \frac{1}{-p_{0}^{2}+\mbox{\boldmath $p$}^{2}+m^{2}
     +\frac{\lambda}{2}
     \bar{\phi}^{2}+\Pi}\frac{1}{e^{p_{0}/T}-1}\nonumber \\
     &&+\frac{1}{4\pi i}\int^{+i\infty}_{-i\infty}dp_{0}
     \int\frac{d^{3}\mbox{\boldmath $p$}}{(2\pi)^{3}}
     \frac{1}{-p_{0}^{2}+\mbox{\boldmath $p$}^{2}+m^{2}
     +\frac{\lambda}{2}\bar{\phi}^{2}+\Pi}\nonumber \\
     &&+(Z_{m}Z_{\phi}-1)\biggl[\frac{1}{2}\bar{\phi}^{2}
     \nonumber \\
     &&+\frac{1}{2\pi i}
     \int^{+i\infty +\epsilon}_{-i\infty +\epsilon}dp_{0}
     \int\frac{d^{3}\mbox{\boldmath $p$}}{(2\pi)^{3}}
     \frac{1}{-p_{0}^{2}+\mbox{\boldmath $p$}^{2}+m^{2}
     +\frac{\lambda}{2}
     \bar{\phi}^{2}+\Pi}\frac{1}{e^{p_{0}/T}-1}\nonumber \\
     &&+\frac{1}{4\pi i}\int^{+i\infty}_{-i\infty}dp_{0}
     \int\frac{d^{3}\mbox{\boldmath $p$}}{(2\pi)^{3}}
     \frac{1}{-p_{0}^{2}+\mbox{\boldmath $p$}^{2}+m^{2}
     +\frac{\lambda}{2}\bar{\phi}^{2}+\Pi}\biggr].
\label{delvsep}
\end{eqnarray}

\end{document}

%% file: epot1_eepic.tex
% GNUPLOT: LaTeX picture using EEPIC macros
\setlength{\unitlength}{0.240900pt}
\begin{picture}(900,900)(0,0)
%\tenrm
\thinlines \drawline[-50](220,113)(220,877)
%\thicklines \path(220,113)(240,113)
%\thicklines \path(836,113)(816,113)
%\put(198,113){\makebox(0,0)[r]{-0.005}}
\thicklines \path(220,189)(240,189)
\thicklines \path(836,189)(816,189)
\put(198,189){\makebox(0,0)[r]{-0.004}}
%\thicklines \path(220,266)(240,266)
%\thicklines \path(836,266)(816,266)
%\put(198,266){\makebox(0,0)[r]{-0.003}}
\thicklines \path(220,342)(240,342)
\thicklines \path(836,342)(816,342)
\put(198,342){\makebox(0,0)[r]{-0.002}}
%\thicklines \path(220,419)(240,419)
%\thicklines \path(836,419)(816,419)
%\put(198,419){\makebox(0,0)[r]{-0.001}}
\thicklines \path(220,495)(240,495)
\thicklines \path(836,495)(816,495)
\put(198,495){\makebox(0,0)[r]{0}}
%\thicklines \path(220,571)(240,571)
%\thicklines \path(836,571)(816,571)
%\put(198,571){\makebox(0,0)[r]{0.001}}
\thicklines \path(220,648)(240,648)
\thicklines \path(836,648)(816,648)
\put(198,648){\makebox(0,0)[r]{0.002}}
%\thicklines \path(220,724)(240,724)
%\thicklines \path(836,724)(816,724)
%\put(198,724){\makebox(0,0)[r]{0.003}}
\thicklines \path(220,801)(240,801)
\thicklines \path(836,801)(816,801)
\put(198,801){\makebox(0,0)[r]{0.004}}
%\thicklines \path(220,877)(240,877)
%\thicklines \path(836,877)(816,877)
%\put(198,877){\makebox(0,0)[r]{0.005}}
\thicklines \path(220,113)(220,133)
\thicklines \path(220,877)(220,857)
\put(220,68){\makebox(0,0){0}}
\thicklines \path(343,113)(343,133)
\thicklines \path(343,877)(343,857)
\put(343,68){\makebox(0,0){0.2}}
\thicklines \path(466,113)(466,133)
\thicklines \path(466,877)(466,857)
\put(466,68){\makebox(0,0){0.4}}
\thicklines \path(590,113)(590,133)
\thicklines \path(590,877)(590,857)
\put(590,68){\makebox(0,0){0.6}}
\thicklines \path(713,113)(713,133)
\thicklines \path(713,877)(713,857)
\put(713,68){\makebox(0,0){0.8}}
\thicklines \path(836,113)(836,133)
\thicklines \path(836,877)(836,857)
\put(836,68){\makebox(0,0){1}}
\thicklines \path(220,113)(836,113)(836,877)(220,877)(220,113)
\put(10,495){\makebox(0,0)[l]{\shortstack{$\displaystyle \Re \frac{V(\phi)}{\mu^{4}}$}}}
\put(528,23){\makebox(0,0){$\phi/\mu$}}
\put(282,648){\makebox(0,0)[l]{$T=5.1\mu$}}
\put(418,470){\makebox(0,0)[l]{$T=5.05\mu$}}
\put(528,357){\makebox(0,0)[l]{$T=5.0\mu$}}
\put(269,266){\makebox(0,0)[l]{$T=4.95\mu$}}
\thinlines \path(220,495)(220,495)(307,503)(394,535)(481,610)(568,770)(598,877)
\thinlines \path(220,495)(220,495)(307,495)(394,496)(481,516)(568,594)(656,796)(672,877)
\thinlines \path(220,495)(220,495)(307,483)(394,453)(481,417)(568,403)(656,475)(743,729)(766,877)
\thinlines \path(220,495)(220,495)(307,469)(394,398)(481,299)(568,199)(656,149)(743,242)(830,606)(836,663)
\end{picture}

%% file: epotper1_eepic.tex
% GNUPLOT: LaTeX picture using EEPIC macros
\setlength{\unitlength}{0.240900pt}
\begin{picture}(900,900)(0,0)
%\tenrm
\thinlines \drawline[-50](220,113)(220,877)
\thicklines \path(220,189)(240,189)
\thicklines \path(836,189)(816,189)
\put(198,189){\makebox(0,0)[r]{-0.004}}
\thicklines \path(220,342)(240,342)
\thicklines \path(836,342)(816,342)
\put(198,342){\makebox(0,0)[r]{-0.002}}
\thicklines \path(220,495)(240,495)
\thicklines \path(836,495)(816,495)
\put(198,495){\makebox(0,0)[r]{0}}
\thicklines \path(220,648)(240,648)
\thicklines \path(836,648)(816,648)
\put(198,648){\makebox(0,0)[r]{0.002}}
\thicklines \path(220,801)(240,801)
\thicklines \path(836,801)(816,801)
\put(198,801){\makebox(0,0)[r]{0.004}}
\thicklines \path(220,113)(220,133)
\thicklines \path(220,877)(220,857)
\put(220,68){\makebox(0,0){0}}
\thicklines \path(343,113)(343,133)
\thicklines \path(343,877)(343,857)
\put(343,68){\makebox(0,0){0.2}}
\thicklines \path(466,113)(466,133)
\thicklines \path(466,877)(466,857)
\put(466,68){\makebox(0,0){0.4}}
\thicklines \path(590,113)(590,133)
\thicklines \path(590,877)(590,857)
\put(590,68){\makebox(0,0){0.6}}
\thicklines \path(713,113)(713,133)
\thicklines \path(713,877)(713,857)
\put(713,68){\makebox(0,0){0.8}}
\thicklines \path(836,113)(836,133)
\thicklines \path(836,877)(836,857)
\put(836,68){\makebox(0,0){1}}
\thicklines \path(220,113)(836,113)(836,877)(220,877)(220,113)
\put(10,495){\makebox(0,0)[l]{\shortstack{$\displaystyle \Re \frac{V(\phi)}{\mu^{4}}$}}}
\put(528,23){\makebox(0,0){$\phi/\mu$}}
\put(251,709){\makebox(0,0)[l]{$T=5.1\mu$}}
\put(454,556){\makebox(0,0)[l]{$T=5.05\mu$}}
\put(454,472){\makebox(0,0)[l]{$T=5.0\mu$}}
\put(528,312){\makebox(0,0)[l]{$T=4.95\mu$}}
\thinlines \path(220,495)(220,495)(221,495)(222,495)(222,495)(223,495)(224,495)(225,495)(226,495)(228,495)(230,495)(233,495)(236,496)(239,496)(246,497)(252,498)(259,499)(271,502)(284,507)(297,512)(323,525)(348,542)(374,563)(400,588)(425,618)(451,654)(477,696)(502,745)(528,803)(554,870)(556,877)
\thinlines \path(220,495)(220,495)(221,495)(222,495)(222,495)(223,495)(224,495)(225,495)(226,495)(228,495)(230,495)(233,495)(236,495)(239,496)(246,496)(252,496)(259,497)(271,499)(284,501)(297,504)(323,510)(348,519)(374,529)(400,542)(425,557)(451,576)(477,598)(502,626)(528,659)(554,700)(579,749)(605,810)(629,877)
\thinlines \path(220,495)(220,495)(221,495)(222,495)(222,495)(223,495)(224,495)(225,495)(226,495)(228,495)(230,495)(233,495)(239,495)(246,495)(259,495)(271,496)(297,496)(323,497)(348,498)(374,498)(400,499)(413,500)(425,500)(438,501)(451,502)(464,504)(477,506)(490,508)(502,512)(515,516)(528,521)(541,528)(554,536)(579,556)(605,586)(631,625)(656,677)(682,745)(708,830)(719,877)
\thinlines \path(220,495)(220,495)(221,495)(222,495)(222,495)(223,495)(224,495)(225,495)(226,495)(228,495)(230,495)(233,495)(236,495)(239,495)(246,495)(252,494)(259,494)(271,493)(284,492)(297,490)(310,488)(323,486)(348,480)(374,471)(400,461)(425,449)(451,435)(477,420)(502,405)(528,392)(541,385)(554,380)(567,376)(573,374)(579,373)(583,372)(586,372)(589,372)(591,371)(592,371)(593,371)(594,371)(595,371)(595,371)(596,371)(597,371)(598,371)(599,371)(599,371)(600,371)(602,371)
\thinlines \path(602,371)(603,371)(603,371)(605,371)(608,372)(611,372)(615,373)(618,374)(624,375)(631,378)(644,385)(656,394)(669,407)(682,424)(708,469)(733,532)(759,617)(785,728)(810,867)(812,877)
\end{picture}

%% file: epotper1p_eepic.tex
% GNUPLOT: LaTeX picture using EEPIC macros
\setlength{\unitlength}{0.240900pt}
\begin{picture}(900,900)(0,0)
\thinlines \drawline[-50](220,113)(220,877)
\thicklines \path(220,113)(240,113)
\thicklines \path(836,113)(816,113)
\put(198,113){\makebox(0,0)[r]{$-1\times 10^{-5}$}}
\thicklines \path(220,260)(240,260)
\thicklines \path(836,260)(816,260)
\put(198,260){\makebox(0,0)[r]{$-5\times 10^{-6}$}}
\thicklines \path(220,407)(240,407)
\thicklines \path(836,407)(816,407)
\put(198,407){\makebox(0,0)[r]{0}}
\thicklines \path(220,554)(240,554)
\thicklines \path(836,554)(816,554)
\put(198,554){\makebox(0,0)[r]{$5\times 10^{-6}$}}
\thicklines \path(220,701)(240,701)
\thicklines \path(836,701)(816,701)
\put(198,701){\makebox(0,0)[r]{$1\times 10^{-5}$}}
\thicklines \path(220,848)(240,848)
\thicklines \path(836,848)(816,848)
\put(198,848){\makebox(0,0)[r]{$1.5\times 10^{-5}$}}
\thicklines \path(220,113)(220,133)
\thicklines \path(220,877)(220,857)
\put(220,68){\makebox(0,0){0}}
\thicklines \path(374,113)(374,133)
\thicklines \path(374,877)(374,857)
\put(374,68){\makebox(0,0){0.1}}
\thicklines \path(528,113)(528,133)
\thicklines \path(528,877)(528,857)
\put(528,68){\makebox(0,0){0.2}}
\thicklines \path(682,113)(682,133)
\thicklines \path(682,877)(682,857)
\put(682,68){\makebox(0,0){0.3}}
\thicklines \path(836,113)(836,133)
\thicklines \path(836,877)(836,857)
\put(836,68){\makebox(0,0){0.4}}
\thicklines \path(220,113)(836,113)(836,877)(220,877)(220,113)
\put(10,407){\makebox(0,0)[l]{\shortstack{$\displaystyle \Re \frac{V(\phi)}{\mu^{4}}$}}}
\put(528,23){\makebox(0,0){$\phi/\mu$}}
\put(430,820){\makebox(0,0)[l]{$T=4.9957\mu$}}
\put(430,400){\makebox(0,0)[l]{$T=4.9952\mu$}}
\put(430,200){\makebox(0,0)[l]{$T=4.9947\mu$}}
\thinlines \path(220,407)(220,407)(221,407)(222,407)(222,407)(223,407)(224,407)(225,407)(226,407)(228,407)(230,408)(233,408)(236,409)(239,410)(246,413)(259,420)(271,429)(297,455)(323,489)(348,527)(374,567)(400,605)(425,637)(438,650)(451,660)(457,664)(464,668)(470,670)(473,671)(475,671)(477,672)(478,672)(480,672)(481,672)(481,672)(482,672)(483,672)(484,672)(485,672)(485,672)(486,672)(487,672)(488,672)(490,672)(491,672)(493,671)(496,671)(499,670)(502,669)(509,665)(515,661)
\thinlines \path(515,661)(528,650)(541,635)(554,615)(579,565)(605,502)(631,430)(656,355)(682,282)(695,250)(708,222)(721,200)(727,192)(730,188)(733,185)(737,183)(738,182)(740,181)(741,181)(742,180)(743,180)(744,180)(745,180)(745,180)(746,180)(747,180)(748,180)(749,180)(749,180)(750,180)(751,180)(753,181)(754,181)(756,182)(759,185)(762,188)(765,192)(772,203)(778,217)(785,236)(798,286)(810,355)(823,446)(836,562)
\thinlines \path(220,407)(220,407)(221,407)(222,407)(222,407)(223,407)(224,407)(225,407)(226,407)(228,407)(230,408)(233,408)(236,409)(239,410)(246,413)(259,421)(271,431)(297,459)(323,496)(348,539)(374,584)(400,628)(425,668)(438,685)(451,700)(464,712)(470,717)(477,721)(483,724)(486,726)(490,727)(493,728)(494,728)(496,728)(498,729)(499,729)(501,729)(502,729)(502,729)(503,729)(504,729)(505,729)(506,729)(506,729)(507,729)(508,729)(509,729)(510,729)(512,728)(515,728)(518,727)
\thinlines \path(518,727)(522,726)(528,723)(534,719)(541,714)(554,702)(579,667)(605,620)(631,566)(656,509)(682,458)(695,436)(701,427)(708,420)(714,414)(717,411)(721,409)(722,409)(724,408)(725,407)(726,407)(727,407)(728,407)(729,407)(729,407)(730,407)(731,407)(732,407)(733,407)(733,407)(735,407)(736,407)(737,407)(738,408)(740,409)(743,411)(746,413)(753,421)(759,431)(765,445)(772,463)(785,509)(798,573)(810,657)(834,877)
\thinlines \path(220,407)(220,407)(221,407)(222,407)(222,407)(223,407)(224,407)(225,407)(226,407)(228,407)(230,408)(233,408)(236,409)(239,411)(246,413)(259,421)(271,433)(297,463)(323,503)(348,550)(374,599)(400,649)(425,696)(451,735)(464,751)(477,765)(483,771)(490,776)(496,780)(502,783)(506,784)(509,785)(512,786)(514,787)(515,787)(517,787)(518,787)(519,788)(520,788)(521,788)(522,788)(522,788)(523,788)(524,788)(525,788)(526,788)(526,788)(527,788)(528,788)(530,787)(531,787)
\thinlines \path(531,787)(534,787)(538,786)(541,785)(547,782)(554,779)(567,769)(579,757)(605,725)(631,687)(656,647)(669,629)(682,614)(688,607)(695,602)(698,600)(701,598)(704,596)(706,596)(708,595)(709,595)(710,595)(711,595)(712,595)(712,595)(713,594)(714,594)(715,594)(716,594)(716,595)(717,595)(718,595)(719,595)(721,595)(722,596)(724,597)(727,598)(730,600)(733,603)(740,611)(746,621)(753,634)(759,650)(772,693)(785,752)(798,828)(804,877)
\end{picture}

%% file: epotper2_eepic.tex
% GNUPLOT: LaTeX picture using EEPIC macros
\setlength{\unitlength}{0.240900pt}
\begin{picture}(900,900)(0,0)
%\tenrm
\thinlines \drawline[-50](220,113)(220,877)
\thicklines \path(220,189)(240,189)
\thicklines \path(836,189)(816,189)
\put(198,189){\makebox(0,0)[r]{-0.004}}
\thicklines \path(220,342)(240,342)
\thicklines \path(836,342)(816,342)
\put(198,342){\makebox(0,0)[r]{-0.002}}
\thicklines \path(220,495)(240,495)
\thicklines \path(836,495)(816,495)
\put(198,495){\makebox(0,0)[r]{0}}
\thicklines \path(220,648)(240,648)
\thicklines \path(836,648)(816,648)
\put(198,648){\makebox(0,0)[r]{0.002}}
\thicklines \path(220,801)(240,801)
\thicklines \path(836,801)(816,801)
\put(198,801){\makebox(0,0)[r]{0.004}}
\thicklines \path(220,113)(220,133)
\thicklines \path(220,877)(220,857)
\put(220,68){\makebox(0,0){0}}
\thicklines \path(343,113)(343,133)
\thicklines \path(343,877)(343,857)
\put(343,68){\makebox(0,0){0.2}}
\thicklines \path(466,113)(466,133)
\thicklines \path(466,877)(466,857)
\put(466,68){\makebox(0,0){0.4}}
\thicklines \path(590,113)(590,133)
\thicklines \path(590,877)(590,857)
\put(590,68){\makebox(0,0){0.6}}
\thicklines \path(713,113)(713,133)
\thicklines \path(713,877)(713,857)
\put(713,68){\makebox(0,0){0.8}}
\thicklines \path(836,113)(836,133)
\thicklines \path(836,877)(836,857)
\put(836,68){\makebox(0,0){1}}
\thicklines \path(220,113)(836,113)(836,877)(220,877)(220,113)
\put(10,495){\makebox(0,0)[l]{\shortstack{$\displaystyle \Re \frac{V(\phi)}{\mu^{4}}$}}}
\put(528,23){\makebox(0,0){$\phi/\mu$}}
\put(282,648){\makebox(0,0)[l]{$T=5.1\mu$}}
\put(436,449){\makebox(0,0)[l]{$T=5.05\mu$}}
\put(528,312){\makebox(0,0)[l]{$T=5.0\mu$}}
\put(282,266){\makebox(0,0)[l]{$T=4.95\mu$}}
\thinlines \path(220,495)(220,495)(221,495)(222,495)(222,495)(223,495)(224,495)(225,495)(226,495)(228,495)(230,495)(233,495)(236,495)(239,495)(246,496)(252,496)(259,496)(271,497)(284,498)(297,500)(310,502)(323,504)(348,510)(374,519)(400,530)(425,545)(451,564)(477,589)(502,620)(528,659)(554,707)(579,766)(605,838)(617,877)
\thinlines \path(220,495)(220,495)(221,495)(222,495)(222,495)(223,495)(224,495)(225,495)(226,495)(228,495)(230,495)(233,495)(236,495)(239,495)(246,495)(259,494)(271,493)(297,491)(323,489)(348,487)(361,485)(374,484)(380,484)(387,484)(393,483)(396,483)(400,483)(401,483)(403,483)(404,483)(405,483)(406,483)(407,483)(408,483)(408,483)(409,483)(410,483)(411,483)(412,483)(413,483)(414,483)(415,483)(416,483)(419,483)(421,483)(422,483)(425,483)(432,484)(438,484)(445,485)(451,486)
\thinlines \path(451,486)(464,488)(477,491)(490,496)(502,501)(515,508)(528,517)(554,539)(579,569)(605,611)(631,664)(656,733)(682,819)(696,877)
\thinlines \path(220,495)(220,495)(221,495)(222,495)(222,495)(223,495)(224,495)(225,495)(226,495)(228,495)(230,495)(233,495)(236,494)(239,494)(246,494)(259,492)(271,490)(297,483)(323,475)(348,464)(374,452)(400,439)(425,425)(451,411)(477,398)(502,387)(515,383)(522,381)(528,380)(534,378)(541,377)(544,377)(546,377)(547,377)(549,377)(550,376)(551,376)(552,376)(553,376)(554,376)(554,376)(555,376)(556,376)(557,376)(558,376)(558,376)(560,376)(562,377)(563,377)(567,377)(570,377)
\thinlines \path(570,377)(573,378)(579,379)(586,381)(592,384)(605,390)(618,400)(631,412)(644,427)(656,445)(682,493)(708,559)(733,646)(759,756)(781,877)
\thinlines \path(220,495)(220,495)(221,495)(222,495)(222,495)(223,495)(224,495)(225,495)(226,495)(228,495)(230,495)(233,494)(236,494)(239,494)(246,493)(259,490)(271,486)(297,476)(323,462)(348,444)(374,422)(400,398)(425,371)(451,342)(477,311)(502,280)(528,250)(554,222)(579,198)(592,188)(605,180)(611,176)(618,173)(624,171)(627,170)(631,169)(634,168)(635,168)(637,168)(639,167)(640,167)(641,167)(642,167)(643,167)(644,167)(644,167)(645,167)(646,167)(647,167)(648,167)(648,167)
\thinlines \path(648,167)(649,167)(650,167)(652,167)(653,168)(655,168)(656,168)(660,169)(663,170)(669,172)(676,175)(682,180)(695,191)(708,206)(721,226)(733,251)(759,318)(785,409)(810,528)(836,680)
\end{picture}

%% file: epot2_eepic.tex
% GNUPLOT: LaTeX picture using EEPIC macros
\setlength{\unitlength}{0.240900pt}
\begin{picture}(1500,900)(0,0)
%\tenrm
\thinlines \drawline[-50](220,113)(220,877)
\thicklines \path(220,113)(240,113)
\thicklines \path(1436,113)(1416,113)
\put(198,113){\makebox(0,0)[r]{-0.0005}}
%\thicklines \path(220,189)(240,189)
%\thicklines \path(1436,189)(1416,189)
%\put(198,189){\makebox(0,0)[r]{-0.00045}}
\thicklines \path(220,266)(240,266)
\thicklines \path(1436,266)(1416,266)
\put(198,266){\makebox(0,0)[r]{-0.0004}}
%\thicklines \path(220,342)(240,342)
%\thicklines \path(1436,342)(1416,342)
%\put(198,342){\makebox(0,0)[r]{-0.00035}}
\thicklines \path(220,419)(240,419)
\thicklines \path(1436,419)(1416,419)
\put(198,419){\makebox(0,0)[r]{-0.0003}}
%\thicklines \path(220,495)(240,495)
%\thicklines \path(1436,495)(1416,495)
%\put(198,495){\makebox(0,0)[r]{-0.00025}}
\thicklines \path(220,571)(240,571)
\thicklines \path(1436,571)(1416,571)
\put(198,571){\makebox(0,0)[r]{-0.0002}}
%\thicklines \path(220,648)(240,648)
%\thicklines \path(1436,648)(1416,648)
%\put(198,648){\makebox(0,0)[r]{-0.00015}}
\thicklines \path(220,724)(240,724)
\thicklines \path(1436,724)(1416,724)
\put(198,724){\makebox(0,0)[r]{-0.0001}}
%\thicklines \path(220,801)(240,801)
%\thicklines \path(1436,801)(1416,801)
%\put(198,801){\makebox(0,0)[r]{-5e-05}}
\thicklines \path(220,877)(240,877)
\thicklines \path(1436,877)(1416,877)
\put(198,877){\makebox(0,0)[r]{0}}
\thicklines \path(220,113)(220,133)
\thicklines \path(220,877)(220,857)
\put(220,68){\makebox(0,0){0}}
\thicklines \path(463,113)(463,133)
\thicklines \path(463,877)(463,857)
\put(463,68){\makebox(0,0){0.2}}
\thicklines \path(706,113)(706,133)
\thicklines \path(706,877)(706,857)
\put(706,68){\makebox(0,0){0.4}}
\thicklines \path(950,113)(950,133)
\thicklines \path(950,877)(950,857)
\put(950,68){\makebox(0,0){0.6}}
\thicklines \path(1193,113)(1193,133)
\thicklines \path(1193,877)(1193,857)
\put(1193,68){\makebox(0,0){0.8}}
\thicklines \path(1436,113)(1436,133)
\thicklines \path(1436,877)(1436,857)
\put(1436,68){\makebox(0,0){1}}
\thicklines \path(220,113)(1436,113)(1436,877)(220,877)(220,113)
\put(45,945){\makebox(0,0)[l]{\shortstack{$\Im V(\phi)/\mu^{4}$}}}
\put(828,23){\makebox(0,0){$\phi/\mu$}}
\put(281,342){\makebox(0,0)[l]{$T=5.0\mu$}}
\put(804,571){\makebox(0,0)[l]{$T=4.95\mu$}}
\thinlines \path(220,877)(220,877)(392,877)(564,877)(736,877)(908,877)(1080,877)(1252,877)(1424,877)(1436,877)
\thinlines \path(220,332)(220,332)(392,451)(564,707)(736,855)(908,876)(1080,877)(1252,877)(1424,877)(1436,877)
\thinlines \path(592,113)(736,575)(908,829)(1080,873)(1252,877)(1424,877)(1436,877)
\end{picture}

%% file: cri_eepic.tex
% GNUPLOT: LaTeX picture using EEPIC macros
\setlength{\unitlength}{0.240900pt}
\begin{picture}(1500,900)(0,0)
%\tenrm
\thicklines \path(220,113)(240,113)
\thicklines \path(1436,113)(1416,113)
\put(198,113){\makebox(0,0)[r]{0}}
\thicklines \path(220,222)(240,222)
\thicklines \path(1436,222)(1416,222)
\put(198,222){\makebox(0,0)[r]{10}}
\thicklines \path(220,331)(240,331)
\thicklines \path(1436,331)(1416,331)
\put(198,331){\makebox(0,0)[r]{20}}
\thicklines \path(220,440)(240,440)
\thicklines \path(1436,440)(1416,440)
\put(198,440){\makebox(0,0)[r]{30}}
\thicklines \path(220,550)(240,550)
\thicklines \path(1436,550)(1416,550)
\put(198,550){\makebox(0,0)[r]{40}}
\thicklines \path(220,659)(240,659)
\thicklines \path(1436,659)(1416,659)
\put(198,659){\makebox(0,0)[r]{50}}
\thicklines \path(220,768)(240,768)
\thicklines \path(1436,768)(1416,768)
\put(198,768){\makebox(0,0)[r]{60}}
\thicklines \path(220,877)(240,877)
\thicklines \path(1436,877)(1416,877)
\put(198,877){\makebox(0,0)[r]{70}}
\thicklines \path(220,113)(220,133)
\thicklines \path(220,877)(220,857)
\put(220,68){\makebox(0,0){0}}
\thicklines \path(463,113)(463,133)
\thicklines \path(463,877)(463,857)
\put(463,68){\makebox(0,0){0.2}}
\thicklines \path(706,113)(706,133)
\thicklines \path(706,877)(706,857)
\put(706,68){\makebox(0,0){0.4}}
\thicklines \path(950,113)(950,133)
\thicklines \path(950,877)(950,857)
\put(950,68){\makebox(0,0){0.6}}
\thicklines \path(1193,113)(1193,133)
\thicklines \path(1193,877)(1193,857)
\put(1193,68){\makebox(0,0){0.8}}
\thicklines \path(1436,113)(1436,133)
\thicklines \path(1436,877)(1436,857)
\put(1436,68){\makebox(0,0){1}}
\thicklines \path(220,113)(1436,113)(1436,877)(220,877)(220,113)
\put(45,495){\makebox(0,0)[l]{\shortstack{$T_{c}/\mu$}}}
\put(828,23){\makebox(0,0){$\lambda$}}
\put(706,659){\makebox(0,0)[l]{$\phi_{c}=0$}}
\put(281,168){\makebox(0,0)[l]{$\phi_{c}\neq 0$}}
\thinlines \path(1436,168)(1436,168)(828,190)(524,221)(372,266)(342,283)(281,353)(250,453)(235,593)(232,649)(226,871)(226,877)
\end{picture}

%% file: criphi_eepic.tex
% GNUPLOT: LaTeX picture using EEPIC macros
\setlength{\unitlength}{0.240900pt}
\begin{picture}(900,900)(0,0)
%\tenrm
\thicklines \path(220,113)(240,113)
\thicklines \path(836,113)(816,113)
\put(198,113){\makebox(0,0)[r]{0}}
\thicklines \path(220,252)(240,252)
\thicklines \path(836,252)(816,252)
\put(198,252){\makebox(0,0)[r]{0.2}}
\thicklines \path(220,391)(240,391)
\thicklines \path(836,391)(816,391)
\put(198,391){\makebox(0,0)[r]{0.4}}
\thicklines \path(220,530)(240,530)
\thicklines \path(836,530)(816,530)
\put(198,530){\makebox(0,0)[r]{0.6}}
\thicklines \path(220,669)(240,669)
\thicklines \path(836,669)(816,669)
\put(198,669){\makebox(0,0)[r]{0.8}}
\thicklines \path(220,808)(240,808)
\thicklines \path(836,808)(816,808)
\put(198,808){\makebox(0,0)[r]{1}}
%\thicklines \path(220,113)(220,133)
%\thicklines \path(220,877)(220,857)
%\put(220,68){\makebox(0,0){4.75}}
\thicklines \path(288,113)(288,133)
\thicklines \path(288,877)(288,857)
\put(288,68){\makebox(0,0){4.8}}
%\thicklines \path(357,113)(357,133)
%\thicklines \path(357,877)(357,857)
%\put(357,68){\makebox(0,0){4.85}}
\thicklines \path(425,113)(425,133)
\thicklines \path(425,877)(425,857)
\put(425,68){\makebox(0,0){4.9}}
%\thicklines \path(494,113)(494,133)
%\thicklines \path(494,877)(494,857)
%\put(494,68){\makebox(0,0){4.95}}
\thicklines \path(562,113)(562,133)
\thicklines \path(562,877)(562,857)
\put(562,68){\makebox(0,0){5}}
%\thicklines \path(631,113)(631,133)
%\thicklines \path(631,877)(631,857)
%\put(631,68){\makebox(0,0){5.05}}
\thicklines \path(699,113)(699,133)
\thicklines \path(699,877)(699,857)
\put(699,68){\makebox(0,0){5.1}}
%\thicklines \path(768,113)(768,133)
%\thicklines \path(768,877)(768,857)
%\put(768,68){\makebox(0,0){5.15}}
\thicklines \path(836,113)(836,133)
\thicklines \path(836,877)(836,857)
\put(836,68){\makebox(0,0){5.2}}
\thicklines \path(220,113)(836,113)(836,877)(220,877)(220,113)
\put(45,945){\makebox(0,0)[l]{\shortstack{$\phi_{c}/\mu$}}}
\put(528,23){\makebox(0,0){$T/\mu$}}
\thinlines \path(836,113)(836,113)(822,113)(809,113)(795,113)(781,113)(768,113)(754,113)(740,113)(726,113)(713,113)(699,113)(685,113)(672,113)(658,113)(644,113)(631,241)(617,313)(603,365)(590,401)(576,433)(562,461)(549,486)(535,509)(521,530)(507,550)(494,568)(480,586)(466,602)(453,618)(439,633)(425,648)(412,662)(398,676)(384,689)(371,702)(357,714)(343,726)(330,737)(316,749)(302,760)(288,771)(275,781)(261,791)(247,802)(234,811)(220,821)(220,821)
\end{picture}

%% file: crirho_eepic.tex
% GNUPLOT: LaTeX picture using EEPIC macros
\setlength{\unitlength}{0.240900pt}
\begin{picture}(900,900)(0,0)
%\tenrm
\thicklines \path(220,113)(240,113)
\thicklines \path(836,113)(816,113)
\put(198,113){\makebox(0,0)[r]{0}}
\thicklines \path(220,254)(240,254)
\thicklines \path(836,254)(816,254)
\put(198,254){\makebox(0,0)[r]{0.05}}
\thicklines \path(220,396)(240,396)
\thicklines \path(836,396)(816,396)
\put(198,396){\makebox(0,0)[r]{0.1}}
\thicklines \path(220,537)(240,537)
\thicklines \path(836,537)(816,537)
\put(198,537){\makebox(0,0)[r]{0.15}}
\thicklines \path(220,679)(240,679)
\thicklines \path(836,679)(816,679)
\put(198,679){\makebox(0,0)[r]{0.2}}
\thicklines \path(220,820)(240,820)
\thicklines \path(836,820)(816,820)
\put(198,820){\makebox(0,0)[r]{0.25}}
%\thicklines \path(220,113)(220,133)
%\thicklines \path(220,877)(220,857)
%\put(220,68){\makebox(0,0){4.75}}
\thicklines \path(288,113)(288,133)
\thicklines \path(288,877)(288,857)
\put(288,68){\makebox(0,0){4.8}}
%\thicklines \path(357,113)(357,133)
%\thicklines \path(357,877)(357,857)
%\put(357,68){\makebox(0,0){4.85}}
\thicklines \path(425,113)(425,133)
\thicklines \path(425,877)(425,857)
\put(425,68){\makebox(0,0){4.9}}
%\thicklines \path(494,113)(494,133)
%\thicklines \path(494,877)(494,857)
%\put(494,68){\makebox(0,0){4.95}}
\thicklines \path(562,113)(562,133)
\thicklines \path(562,877)(562,857)
\put(562,68){\makebox(0,0){5}}
%\thicklines \path(631,113)(631,133)
%\thicklines \path(631,877)(631,857)
%\put(631,68){\makebox(0,0){5.05}}
\thicklines \path(699,113)(699,133)
\thicklines \path(699,877)(699,857)
\put(699,68){\makebox(0,0){5.1}}
%\thicklines \path(768,113)(768,133)
%\thicklines \path(768,877)(768,857)
%\put(768,68){\makebox(0,0){5.15}}
\thicklines \path(836,113)(836,133)
\thicklines \path(836,877)(836,857)
\put(836,68){\makebox(0,0){5.2}}
\thicklines \path(220,113)(836,113)(836,877)(220,877)(220,113)
\put(45,945){\makebox(0,0)[l]{\shortstack{$\rho/\mu$}}}
\put(528,23){\makebox(0,0){$T/\mu$}}
\thinlines \path(836,225)(836,225)(822,216)(809,207)(795,199)(781,190)(768,182)(754,174)(740,166)(726,158)(713,150)(699,142)(685,135)(672,128)(658,121)(644,116)(631,119)(617,133)(603,155)(590,175)(576,194)(562,215)(549,239)(535,260)(521,282)(507,303)(494,326)(480,349)(466,371)(453,394)(439,416)(425,438)(412,460)(398,482)(384,505)(371,528)(357,550)(343,573)(330,595)(316,617)(302,639)(288,661)(275,684)(261,706)(247,728)(234,751)(220,773)(220,773)
\end{picture}

%% file: criv_eepic.tex
% GNUPLOT: LaTeX picture using EEPIC macros
\setlength{\unitlength}{0.240900pt}
\begin{picture}(900,900)(0,0)
%\tenrm
\thicklines \path(220,168)(240,168)
\thicklines \path(836,168)(816,168)
\put(198,168){\makebox(0,0)[r]{-0.025}}
\thicklines \path(220,305)(240,305)
\thicklines \path(836,305)(816,305)
\put(198,305){\makebox(0,0)[r]{-0.02}}
\thicklines \path(220,443)(240,443)
\thicklines \path(836,443)(816,443)
\put(198,443){\makebox(0,0)[r]{-0.015}}
\thicklines \path(220,580)(240,580)
\thicklines \path(836,580)(816,580)
\put(198,580){\makebox(0,0)[r]{-0.01}}
\thicklines \path(220,718)(240,718)
\thicklines \path(836,718)(816,718)
\put(198,718){\makebox(0,0)[r]{-0.005}}
\thicklines \path(220,855)(240,855)
\thicklines \path(836,855)(816,855)
\put(198,855){\makebox(0,0)[r]{0}}
%\thicklines \path(220,113)(220,133)
%\thicklines \path(220,877)(220,857)
%\put(220,68){\makebox(0,0){4.75}}
\thicklines \path(288,113)(288,133)
\thicklines \path(288,877)(288,857)
\put(288,68){\makebox(0,0){4.8}}
%\thicklines \path(357,113)(357,133)
%\thicklines \path(357,877)(357,857)
%\put(357,68){\makebox(0,0){4.85}}
\thicklines \path(425,113)(425,133)
\thicklines \path(425,877)(425,857)
\put(425,68){\makebox(0,0){4.9}}
%\thicklines \path(494,113)(494,133)
%\thicklines \path(494,877)(494,857)
%\put(494,68){\makebox(0,0){4.95}}
\thicklines \path(562,113)(562,133)
\thicklines \path(562,877)(562,857)
\put(562,68){\makebox(0,0){5}}
%\thicklines \path(631,113)(631,133)
%\thicklines \path(631,877)(631,857)
%\put(631,68){\makebox(0,0){5.05}}
\thicklines \path(699,113)(699,133)
\thicklines \path(699,877)(699,857)
\put(699,68){\makebox(0,0){5.1}}
%\thicklines \path(768,113)(768,133)
%\thicklines \path(768,877)(768,857)
%\put(768,68){\makebox(0,0){5.15}}
\thicklines \path(836,113)(836,133)
\thicklines \path(836,877)(836,857)
\put(836,68){\makebox(0,0){5.2}}
\thicklines \path(220,113)(836,113)(836,877)(220,877)(220,113)
\put(45,945){\makebox(0,0)[l]{\shortstack{$V(\phi_{c})/\mu^{4}$}}}
\put(528,23){\makebox(0,0){$T/\mu$}}
\thinlines \path(836,855)(836,855)(822,855)(809,855)(795,855)(781,855)(768,855)(754,855)(740,855)(726,855)(713,855)(699,855)(685,855)(672,855)(658,855)(644,855)(631,855)(617,854)(603,850)(590,846)(576,839)(562,831)(549,821)(535,810)(521,797)(507,783)(494,767)(480,750)(466,730)(453,710)(439,688)(425,665)(412,640)(398,614)(384,586)(371,557)(357,527)(343,495)(330,462)(316,428)(302,392)(288,355)(275,317)(261,277)(247,237)(234,194)(220,152)(220,152)
\end{picture}